\RequirePackage{lineno}
\documentclass[reprint, twocolumn,aps, prl,floatfix,superscriptaddress]{revtex4}

\usepackage{graphicx}         
\usepackage{amsmath,amssymb}  
\usepackage{bm}               
\usepackage{hyperref}         
\usepackage{xcolor}           
\usepackage{float}            
\usepackage{verbatim}         
\usepackage{dcolumn}          
\usepackage[T1]{fontenc}      

\begin{document}
\title{
Evidence of Spin-Interference Effects in Exclusive $J/\psi \to e^+e^-$ Photoproduction in Ultraperipheral Heavy-Ion Collisions
}


\affiliation{Academia Sinica, Nankang, 115}
\affiliation{Abilene Christian University, Abilene, Texas   79699}
\affiliation{AGH University of Krakow, FPACS, Cracow 30-059, Poland}
\affiliation{Argonne National Laboratory, Argonne, Illinois 60439}
\affiliation{American University in Cairo, New Cairo 11835, Egypt}
\affiliation{Ball State University, Muncie, Indiana, 47306}
\affiliation{Brookhaven National Laboratory, Upton, New York 11973}
\affiliation{University of Calabria \& INFN-Cosenza, Rende 87036, Italy}
\affiliation{University of California, Berkeley, California 94720}
\affiliation{University of California, Davis, California 95616}
\affiliation{University of California, Los Angeles, California 90095}
\affiliation{University of California, Riverside, California 92521}
\affiliation{Central China Normal University, Wuhan, Hubei 430079 }
\affiliation{University of Illinois at Chicago, Chicago, Illinois 60607}
\affiliation{Chongqing University, Chongqing, 401331}
\affiliation{Creighton University, Omaha, Nebraska 68178}
\affiliation{Czech Technical University in Prague, FNSPE, Prague 115 19, Czech Republic}
\affiliation{Technische Universit\"at Darmstadt, Darmstadt 64289, Germany}
\affiliation{National Institute of Technology Durgapur, Durgapur - 713209, India}
\affiliation{ELTE E\"otv\"os Lor\'and University, Budapest, Hungary H-1117}
\affiliation{Frankfurt Institute for Advanced Studies FIAS, Frankfurt 60438, Germany}
\affiliation{Fudan University, Shanghai, 200433 }
\affiliation{Guangxi Normal University, Guilin, 541004}
\affiliation{University of Heidelberg, Heidelberg 69120, Germany }
\affiliation{University of Houston, Houston, Texas 77204}
\affiliation{Huzhou University, Huzhou, Zhejiang  313000}
\affiliation{Indian Institute of Science Education and Research (IISER), Berhampur 760010 , India}
\affiliation{Indian Institute of Science Education and Research (IISER) Tirupati, Tirupati 517507, India}
\affiliation{Indian Institute Technology, Patna, Bihar 801106, India}
\affiliation{Indiana University, Bloomington, Indiana 47408}
\affiliation{Institute of Modern Physics, Chinese Academy of Sciences, Lanzhou, Gansu 730000 }
\affiliation{University of Jammu, Jammu 180001, India}
\affiliation{Kent State University, Kent, Ohio 44242}
\affiliation{University of Kentucky, Lexington, Kentucky 40506-0055}
\affiliation{Lanzhou University, Lanzhou, 730000}
\affiliation{Lawrence Berkeley National Laboratory, Berkeley, California 94720}
\affiliation{Lehigh University, Bethlehem, Pennsylvania 18015}
\affiliation{Lovely Professional University, Jalandhar - Delhi G.T. Road, Pagwara, Panjab, 144411, India}
\affiliation{Max-Planck-Institut f\"ur Physik, Munich 80805, Germany}
\affiliation{Michigan State University, East Lansing, Michigan 48824}
\affiliation{National Institute of Science Education and Research, HBNI, Jatni 752050, India}
\affiliation{National Cheng Kung University, Tainan 70101 }
\affiliation{Nuclear Physics Institute of the CAS, Rez 250 68, Czech Republic}
\affiliation{The Ohio State University, Columbus, Ohio 43210}
\affiliation{Panjab University, Chandigarh 160014, India}
\affiliation{Purdue University, West Lafayette, Indiana 47907}
\affiliation{Rice University, Houston, Texas 77251}
\affiliation{Rutgers University, Piscataway, New Jersey 08854}
\affiliation{University of Science and Technology of China, Hefei, Anhui 230026}
\affiliation{South China Normal University, Guangzhou, Guangdong 510631}
\affiliation{Sejong University, Seoul, 05006, Korea, Republic Of}
\affiliation{Shandong University, Qingdao, Shandong 266237}
\affiliation{Shanghai Institute of Applied Physics, Chinese Academy of Sciences, Shanghai 201800}
\affiliation{Southern Connecticut State University, New Haven, Connecticut 06515}
\affiliation{State University of New York, Stony Brook, New York 11794}
\affiliation{Instituto de Alta Investigaci\'on, Universidad de Tarapac\'a, Arica 1000000, Chile}
\affiliation{Temple University, Philadelphia, Pennsylvania 19122}
\affiliation{Texas A\&M University, College Station, Texas 77843}
\affiliation{Texas Southern University, Houston, Texas, 77004}
\affiliation{University of Texas, Austin, Texas 78712}
\affiliation{Tsinghua University, Beijing 100084}
\affiliation{University of Tsukuba, Tsukuba, Ibaraki 305-8571, Japan}
\affiliation{University of Chinese Academy of Sciences, Beijing, 101408}
\affiliation{United States Naval Academy, Annapolis, Maryland 21402}
\affiliation{Valparaiso University, Valparaiso, Indiana 46383}
\affiliation{Variable Energy Cyclotron Centre, Kolkata 700064, India}
\affiliation{Warsaw University of Technology, Warsaw 00-661, Poland}
\affiliation{Wayne State University, Detroit, Michigan 48201}
\affiliation{Wuhan University of Science and Technology, Wuhan, Hubei 430065}
\affiliation{Yale University, New Haven, Connecticut 06520}

\author{B.~E.~Aboona}\affiliation{Texas A\&M University, College Station, Texas 77843}
\author{J.~Adam}\affiliation{Czech Technical University in Prague, FNSPE, Prague 115 19, Czech Republic}
\author{L.~Adamczyk}\affiliation{AGH University of Krakow, FPACS, Cracow 30-059, Poland}
\author{I.~Aggarwal}\affiliation{Panjab University, Chandigarh 160014, India}
\author{M.~M.~Aggarwal}\affiliation{Panjab University, Chandigarh 160014, India}
\author{Z.~Ahammed}\affiliation{Variable Energy Cyclotron Centre, Kolkata 700064, India}
\author{A.~K.~Alshammri}\affiliation{Kent State University, Kent, Ohio 44242}
\author{E.~C.~Aschenauer}\affiliation{Brookhaven National Laboratory, Upton, New York 11973}
\author{S.~Aslam}\affiliation{Fudan University, Shanghai, 200433 }
\author{J.~Atchison}\affiliation{Abilene Christian University, Abilene, Texas   79699}
% ...

\collaboration{STAR Collaboration}\noaffiliation


\begin{abstract}


We report the first evidence of spin interference in exclusive $J/\psi \to e^+e^-$ photoproduction in ultraperipheral heavy-ion collisions at STAR at $\sqrt{s_{NN}} = 200$~GeV. In Au+Au collisions, a negative $\cos(2\phi)$ modulation is found for $p_T < 120$~MeV/$c$ with a significance of $3.2\sigma$, while the isobar data (Ru+Ru, Zr+Zr) show a consistent negative modulation with a significance of $1.9\sigma$, opposite in sign to that in $\rho^{0}\!\to\!\pi^+\pi^-$ photoproduction. This establishes for the first time that the interference sign is controlled by the spin structure of the final-state daughters, resolving the ambiguity present in the all-boson $\rho^0$ channel. The compact $J/\psi$ probes gluon distributions at perturbative scales, resulting in a weaker modulation and providing stringent constraints on Color Glass Condensate calculations. These findings demonstrate that spin-dependent interference in heavy vector mesons provides a new, experimentally accessible handle on gluon structure beyond traditional cross-section measurements.

\end{abstract}



\maketitle


Photon-induced processes can reveal the fundamental structures in nature. From crystallography to the cosmic microwave background, the ability of photons to probe hidden structures has shaped our understanding of the universe. In nuclear and particle physics, this idea takes on a new form: by accelerating electrons or ions, a high-energy collider creates intense electromagnetic fields that act like beams of photons. In electron–proton and electron–ion collisions, most famously studied at HERA and envisioned at the future Electron–Ion Collider (EIC), the photon is emitted by the electron and probes the internal structure of the target~\cite{AbdulKhalek:2021gbh}.  Without an $e+$A collider, similar measurements can be performed in ultraperipheral heavy ion collisions (UPCs) where the ions do not interact hadronically, albeit with less control over initial kinematics.
Ultrarelativistic ions generate strong electromagnetic fields due to their large charge and Lorentz boost, which can be treated in the Weizsäcker–Williams (WW) approximation as a flux of quasireal photons
~\cite{Bertulani:2005ru,Klein:2019qfb}. When two ions pass without colliding, a photon from the projectile may fluctuate into a quark-antiquark pair and interact with the gluon field of the target, producing short-lived particles that carry the quantum numbers of the photon and encode the spatial and momentum structure of the gluon field in the target nucleus.

The process is known as exclusive photoproduction, $\textrm{e.g.}$,\ $\gamma^* + A \to V + A$, where $V$ is a vector meson such as $\rho^0$, $\phi$, or $J/\psi$, and $\gamma^*$ is a quasireal photon~\cite{Bertulani:2005ru}.   
In this reaction, the photon may scatter off the gluon field of the target nucleus, either coherently, involving the entire nucleus, or incoherently, on localized substructures. The momentum transfer ($t$) distribution of the produced vector meson ($d\sigma/dt$) is conjugate to the target's spatial gluon density, encoding both the overall nuclear geometry in coherent events and small-scale fluctuations in incoherent processes~\cite{Mantysaari:2016jaz,Mantysaari:2016ykx,Mantysaari:2022sux,Mantysaari:2023prg,Mantysaari:2022ffw}. These processes have been studied extensively in RHIC and the LHC experiments~\cite{PHENIX:2009xtn,STAR:2017enh,STAR:2023nos,STAR:2023vvb,CMS:2016itn,ALICE:2012yye,ALICE:2020ugp,ALICE:2021jnv,ALICE:2021tyx,ALICE:2021gpt,LHCb:2021hoq,CMS:2023snh,ALICE:2023jgu}. Nevertheless, there are several open questions~\cite{Ryskin:1992ui,Klein:1999qj,Munier:2001nr,Guzey:2016piu} regarding the gluon density distribution. 

\begin{figure*}[htb]
    \centering
   \includegraphics[width=0.9\textwidth]{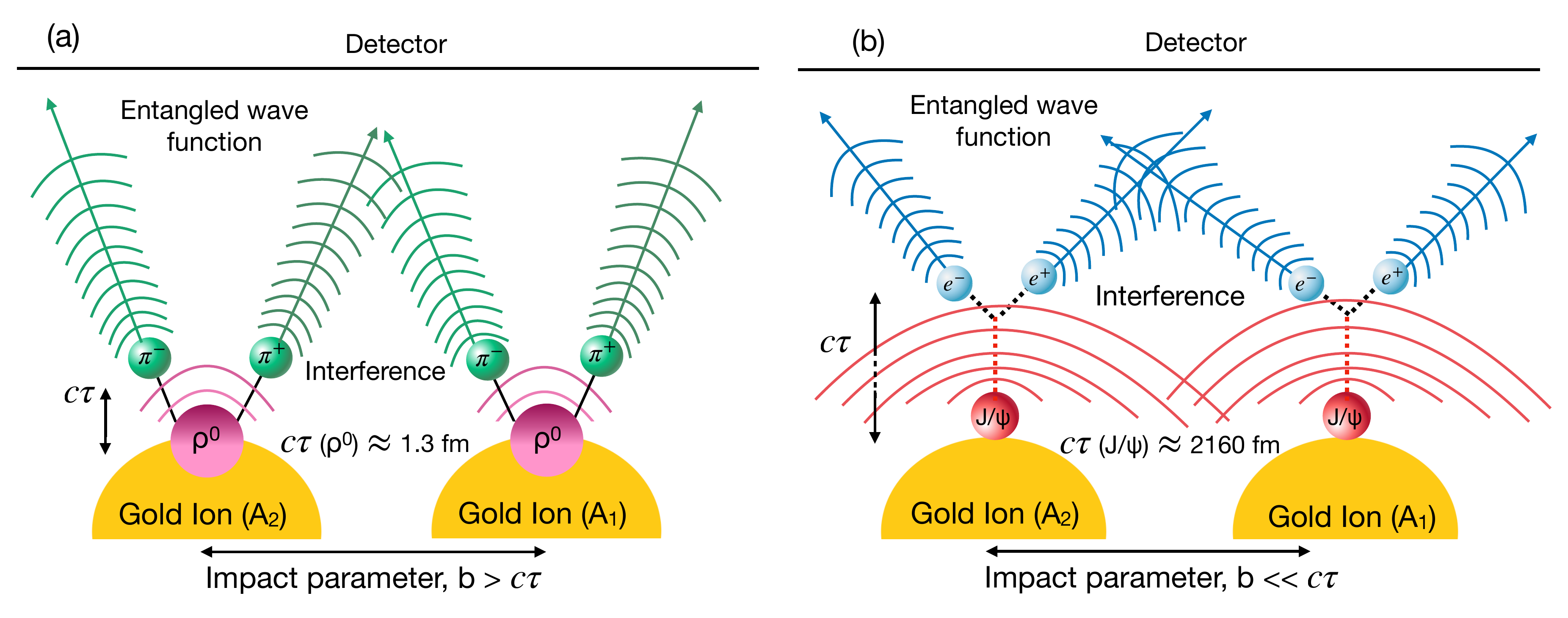}
    \caption{
    Quantum interference in coherent vector meson production. (a) For $\rho^{0}\rightarrow\pi^{+}\pi^{-}$, short lifetime ($c\tau \approx 1.3$ fm $< b$) prevents parent wave function overlap, leaving interference to be carried by entangled spin-0 pion daughters that probe parent spin via orbital angular momentum, yielding a positive modulation. (b) For $J/\psi\rightarrow e^{+}e^{-}$, long lifetime ($c\tau \approx 2160$ fm $\gg b$) enables direct amplitude overlap. Fermionic (spin-1/2) $e^{+}e^{-}$ daughters impose a sign flip, resulting in negative modulation. The contrast between these patterns identifies the final-state spin structure as the driver of the interference. 
}
    \label{cartoon2}
\end{figure*}

Notably, photons in UPCs are linearly polarized in the transverse plane, as demonstrated
by the STAR collaboration in a previous publication~\cite{STAR:2019wlg}. In vector meson photoproduction, the spin-1 photon transfers its polarization to the produced meson, aligning their spin directions. When the vector meson decays, its daughters’ momenta tend to align with the parent spin due to conservation of angular momentum, resulting in an anisotropic azimuthal distribution:
\begin{equation}
    \frac{dN}{d\phi} \propto 1 + a_2 \cos(2\phi)
\end{equation}
Here, $\phi$ is the angle of vector meson momentum ($\vec{p}$) relative to the parent spin direction or photon polarization axis, and $a_2$ is the modulation amplitude. In UPCs, photon polarization is approximately aligned with the impact parameter vector ($\vec{b}$), which is random from event to event and not directly accessible experimentally. Averaging over many events with random polarization orientations would wash out any $\cos(2\phi)$ asymmetry. 

In coherent vector-meson production, where both nuclei remain intact, it is unknown which one emits the photon and which acts as the target. This two-path ambiguity 
contributes at the amplitude level; and a phase difference, $e^{i\vec{p}\cdot\vec{b}}$, between the wave functions in these two paths~\cite{Zha:2020cst,Zha:2018jin,STAR:2022wfe} leads to quantum interference~\cite{Klein:1999gv,Xing:2020hwh}. Hence the $\cos(2\phi)$ asymmetry survives~\cite{STAR:2022wfe}, analogous to a double-slit interference pattern in optics~\cite{Zha:2020cst}. In experimental measurements of $\phi$, the daughters’ momenta are used as a proxy for the spin of the vector meson or the direction of the photon polarization. Experimentally, $\phi$ is therefore taken to be the angle in the transverse plane between the vector meson's momentum and the relative momentum of its decay daughters~\cite{STAR:2022wfe,ALICE:2024ife}: 
\begin{equation}
    \cos\phi = \frac{(\Vec{p}_{T1}+\Vec{p}_{T2}). (\Vec{p}_{T1}-\Vec{p}_{T2})}{|\Vec{p}_{T1}+\Vec{p}_{T2}||\Vec{p}_{T1}-\Vec{p}_{T2}|}
    \label{eq1}
\end{equation}
where $\Vec{p_T}_1$ and $\Vec{p_T}_2$ are transverse momenta of the daughters. 
Refs.~\cite{Klein:1999gv,Klein:2002gc} argue that short-lived vector mesons decay before amplitudes from the two sources can overlap, preventing direct interference. However, the decay products are emitted in an entangled state, and the interference arises from the unitary evolution and subsequent collapse of the full final-state wavefunction ~\cite{STAR:2022wfe,Brandenburg:2024ksp}. 

A $\cos(2\phi)$ modulation, attributed to spin interference, was observed in exclusive $\rho^0$ production in ultra-peripheral Au+Au ~\cite{STAR:2022wfe} and also in Pb+Pb collisions~\cite{ALICE:2024ife}.
%
The interference arises from the overlap of two spatially separated $\rho^0$ wave functions produced coherently in the two nuclei, which are tens of femtometers apart (Fig.~\ref{cartoon2} (a)), far exceeding the $\approx 1$ fm lifetime travel distance of the $\rho^0$~\cite{STAR:2022wfe}. Consequently, the observed modulation must be transmitted through the entangled $\pi^+\pi^-$ daughters~\cite{Brandenburg:2024ksp}. Because the $\cos(2\phi)$ modulation is coupled to the diffractive $|t|$ distribution, it provides complementary orientation-dependent information beyond the angle-integrated cross section and should therefore be accounted for in extracting the transverse spatial gluon distribution, since neglecting it can bias the inferred transverse size and gluon density~\cite{STAR:2022wfe,Mantysaari:2023prg}.

%

However, while these effects are critical for tomography, the $\rho^0$ measurements leave key questions unresolved, due to fundamental limitations of the meson and its decay channel.
Since $\pi^+\pi^-$ daughters are spin-0 particles, the parent- and final-state contributions are experimentally indistinguishable, leaving no testable signature to confirm the specific - or indeed any - role of the daughters. 
%
Beyond such interpretational challenges, the $\rho^0$ also faces limitations as a probe of gluon distributions. Its low mass and large transverse dipole size make it less sensitive to small-$x$ evolution, and obscures its suitability for studying saturation effects~\cite{Brandenburg:2024ksp}. 
%
In contrast, $J/\psi \to e^+e^-$ photoproduction provides a cleaner probe that resolves these issues. Its long lifetime ($c\tau \approx 2160$~fm) allows the quantum state to evolve and interfere well beyond the nuclear volume (Fig.~\ref{cartoon2}~(b)). 
The spin-1/2 $e^+e^-$ daughters imposes a negative sign on the $\cos(2\phi)$ modulation. This qualitative sign flip compared to the spin-0 daughters provides a clear, testable fingerprint of the daughters’ role in spin interference, and removes the ambiguity present in the $\rho^0$ channel. 
%
%
Additionally, its mass justifies perturbative treatment without suppressing sensitivity to the saturation regime~\cite{Brandenburg:2024ksp}. In the kinematic range of this measurement ($|y| < 1$), the $J/\psi$ probes the nuclear gluon distribution at a momentum fraction $x = \frac{M_{V}}{\sqrt{s_{NN}}} e^{\pm y} \approx 0.005\text{--}0.04$. The Color Glass Condensate (CGC) framework is the state-of-the-art approach for describing such saturation effects in dipole-like photoproduction. It has been successfully applied to describe $J/\psi$ photoproduction cross-sections at RHIC energies and provides the only available calculations for the spin-interference signals in these systems~\cite{Mantysaari:2022sux,STAR:2023vvb}.

%
%

In this Letter, we present the first measurement of the spin-interference angular modulation for exclusive $J/\psi$ photoproduction in UPCs, comparing Au+Au to isobaric (Ru+Ru and Zr+Zr) collisions at $\sqrt{s_{NN}} = 200$~GeV. The results are contrasted with the existing $\rho^0$ measurements and compared with CGC model expectations~\cite{Mantysaari:2023prg,Brandenburg:2022jgr}.


We analyze UPC events from Au+Au, Ru+Ru and Zr+Zr collisions at $\sqrt{s_{NN}}=200$ GeV collected with the STAR detector~\cite{STAR:2002eio}, with an integrated luminosity of 13.5, 4 and 3.9 nb$^{-1}$, respectively. Charged particle tracking, including transverse momentum reconstruction, is carried out using the time projection chamber (TPC)~\cite{Anderson:2003ur}, which operates within a 0.5~T solenoidal magnetic field. The TPC extends radially from 50 to 200~cm from the beam axis, covering pseudorapidities $|\eta| < 1.0$ over the azimuthal range $0<\phi<2\pi$. 
It also provides ionization energy loss (\(dE/dx\)) measurements for particle identification. A barrel electromagnetic calorimeter (BEMC)~\cite{STAR:2002ymp} surrounding the TPC is a lead-scintillator sampling calorimeter segmented into 4800 optically isolated towers within \(|\eta| < 1.0\). Positioned radially between the TPC and BEMC, the time-of-flight (TOF) system~\cite{Chen:2024aom} is finely segmented in both \(\eta\) and \(\phi\), offering fast trigger signals for charged particles in the range \(|\eta| < 0.9\). Additionally, two beam-beam counters (BBCs)~\cite{Whitten:2008zz} are located on either side of the STAR detector, covering the range \(3.4 < |\eta| < 5.0\), while two zero-degree calorimeters (ZDCs)~\cite{Xu:2016alq}, placed at \( |\eta| > 6.7\), are used to monitor luminosity while detecting forward neutrons. To trigger the UPC events, we require limited activity in TOF and BEMC, and no signal in the BBCs. The analysis includes events from all neutron emission possibilties  (0n0n, 0nXn, and XnXn) seen via the ZDCs. The analysis aims to select events with exclusive $J/\psi \rightarrow e^+e^-$ production, which requires that there be only two tracks in a single event. For very low-$p_T$ $J/\psi$ candidates, the decay tracks exhibit an approximately back-to-back topology, typically leaving hits in opposite sextants of the BEMC ~\cite{STAR:2023nos,STAR:2023vvb}.

\begin{figure}[htb]
    \centering
   \includegraphics[width=0.47\textwidth]{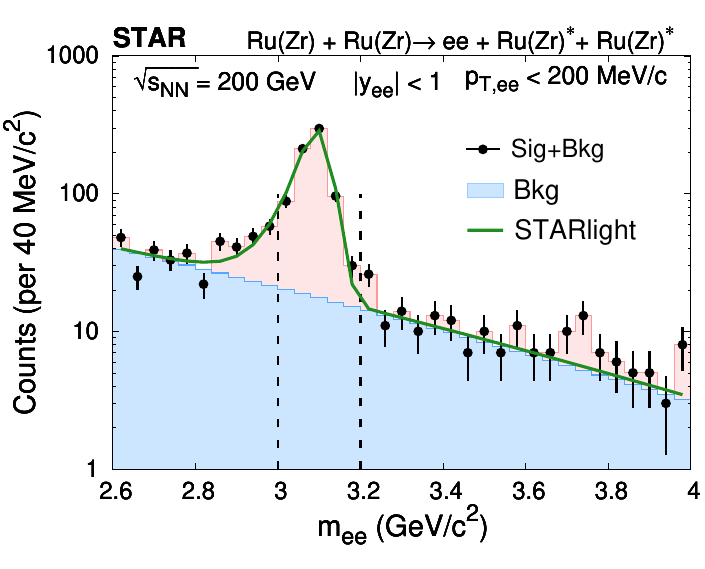}
   \caption{Invariant-mass $m_{ee}$ distribution for $e^+e^-$ pair candidates with $p_{T,ee}<200$ MeV/$c$ and $|y_{ee}|<1$ from Ru+Ru and Zr+Zr UPCs at $\sqrt{s_{NN}}=200$ GeV. The light red and blue regions show the signal-plus-background and background contributions, respectively. The green curve shows the STARlight-based template fit. The vertical dashed lines indicate the chosen $J/\psi$ mass window, $3<m_{ee}<3.2$ GeV/$c^2$.}
    \label{mass}
\end{figure}

We select UPC events which include a pair of tracks with a reconstructed vertex positioned
longitudinally within $\pm$100~cm of the center of the TPC. To ensure adequate momentum resolution, each track must have at least 15 recorded hits out of a possible 45, along with a minimum of 11 hits for an ionization energy loss measurement to maintain reliable \(dE/dx\) resolution. Tracks are also required to point to a BEMC cluster for trigger consistency. The selection of electron pairs relies on track-based \(dE/dx\) values, with pion contamination being the primary background at \(p_T \sim 1.5~\text{GeV}/c\). The variable \(n\sigma_e\) (\(n\sigma_\pi\)) quantifies the deviation of the measured \(dE/dx\) from the expected electron (\(\pi\)) hypothesis, expressed in terms of standard deviations from the theoretical mean. A pair selection variable is defined as: $\chi^2_{ee}=n^2_{\sigma,e1} + n^2_{\sigma,e2}$ for tracks 1 and 2, with a similar formulation applied under the pion pair hypothesis. Electron pair candidates are identified by requiring \(\chi^2_{ee} < 10\), while pion pair contamination is suppressed by enforcing the condition \(\chi^2_{ee} < \chi^2_{\pi\pi}\). These selection criteria are applied independently to opposite-sign (\(+-\)) and like-sign (\(++\), \(--\)) pairs. The like-sign distributions are used to estimate the combinatorial background subsequently subtracted from the opposite-sign pairs to obtain the final signal distributions. The $e^+e^-$ pair invariant-mass ($m_{ee}$) distribution from Ru+Ru and Zr+Zr UPCs at $\sqrt{s_{NN}}=200$ GeV, together with the STARlight-based template fit, is presented in Fig.~\ref{mass}. The background treatment is cross-checked using like-sign, mixed-event, and $\gamma\gamma \rightarrow e^+e^-$ distributions. We also perform the same analysis for Au+Au UPCs at $\sqrt{s_{NN}}=200$ GeV, as presented previously in Refs.~\cite{STAR:2023nos,STAR:2023vvb}.

\begin{figure}[htb]
    \centering
 \includegraphics[width=0.47\textwidth]{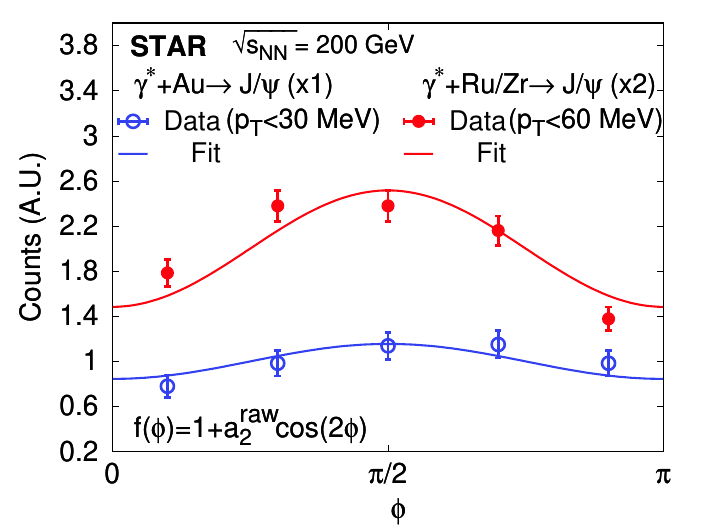}

\caption{Raw azimuthal distribution ($\phi$) of $e^{+}e^{-}$ pairs in the $J/\psi$ mass region ($3 < m_{ee} < 3.2~\mathrm{GeV}/c^{2}$) for the indicated $p_{T}$ ranges.  Solid lines represent fits to $f(\phi) = 1 + a_{2}^{raw}\cos(2\phi)$, with $a_{2}^{raw}$ as fitting parameter. The $a_{2}^{raw}$ values are $-0.155 \pm 0.070$ for Au+Au and $-0.258 \pm 0.076$ for the isobar systems. Au+Au and isobar counts are normalized to 1 and 2, respectively; error bars represent statistical uncertainties.}

    \label{phi}
\end{figure}

\begin{figure*}[htb]
    \centering
   \includegraphics[width=1.0\textwidth]{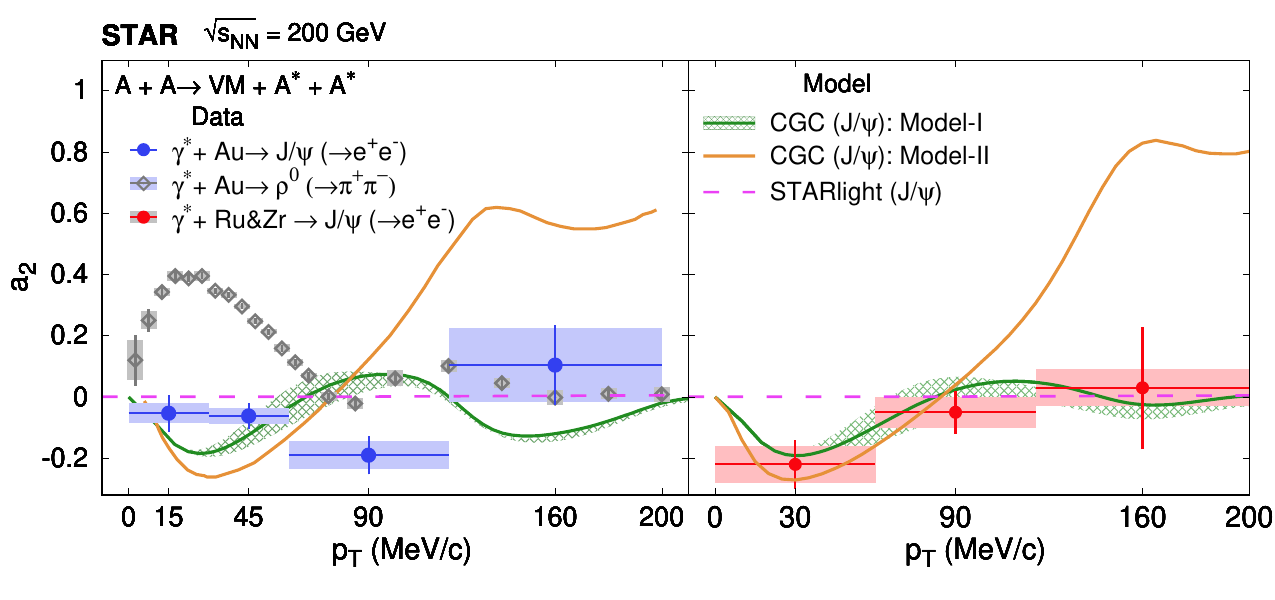}
    \caption{The amplitude of the cos(2$\phi$) modulation, $a_2$, as a function of ${J/\psi}$ $p_T$ from Au+Au (left panel) and Ru+Ru \& Zr+Zr (right panel) UPCs at $\sqrt{s_{NN}}=200$ GeV. The statistical uncertainty on each data point is shown as a  vertical bar, while the systematic uncertainty is shown in the shaded band. The STARlight~\cite{Klein:2016yzr}, CGC (Model-I)~\cite{Mantysaari:2023prg} and CGC (Model-II)~\cite{Brandenburg:2022jgr} calculations are shown with magenta line, green band, and orange line, respectively. STARlight is used here as a spin-independent baseline and not as a full theoretical prediction of the modulation. Model-I incorporates CGC calculations with linear photon polarization and interference effects, while Model-II includes final-state soft-photon radiation.}
    \label{a2}
\end{figure*}

For further analysis, we select the $J/\psi$ decay candidate $e^{+}e^{-}$ pairs in the range $3<m_{ee}<3.2$ GeV/$\textit{c}^{2}$, and the pair $p_T<$ 200 MeV/$\textit{c}$ to measure the $\phi$ observable using Eq.~\ref{eq1}. Figure~\ref{phi} displays the $e^{\pm}$ raw azimuthal correlation functions for the low-$p_{T}$ regions. We selected the smallest $p_T$ bins, as these bins exhibit minimal background contributions and corrections.

%

%
To correct the $a_2$ signal for the combinatorial backgrounds as shown in Fig.~\ref{mass}, we decompose the raw $a_2$ in a two component model: $a_2^{\rm raw} = f\times a_2^{\rm bkg}+(1-f)\times a_2^{\rm sig}$, with $f = \frac{N_{\rm bkg}}{N_{\rm sig}+N_{\rm bkg}}$ being the relative background yield ($N_{\rm bkg}$ is background counts, $N_{\rm sig}+N_{\rm bkg}$ is total counts containing signal+background), obtained from the $e^{+}e^{-}$ pair invariant mass distribution. 
%
The background modulation $a_2^{\rm bkg}$ is estimated from side-band mass intervals ($2.0 < m_{ee} < 2.8$ and $3.2 < m_{ee} < 4.0$~GeV/$c^2$), explicitly excluding the $J/\psi$ signal region ($3.0 < m_{ee} < 3.2$~GeV/$c^2$) and its radiative tail ($2.8 < m_{ee} < 3.0$~GeV/$c^2$) to avoid signal leakage. 
%
%
The $a_2^{\rm sig}$ is background corrected signal modulation which still suffers from detector effects.
The effects of detector acceptance, reconstruction efficiency, and momentum resolution are modeled using STARlight~\cite{Klein:2016yzr} Monte Carlo events embedded into STAR zero-bias events. 
These combined events are then passed through a GEANT3-based~\cite{Brun:1987ma} STAR detector simulation to model the detector response. In this simulation, STARlight includes coherent $J/\psi$, incoherent $J/\psi$ (both with and without nucleon dissociation), as well as QED two-photon ($\gamma\gamma$) processes. 
The simulation specifically incorporates Bremsstrahlung energy loss in the detector material and $p_T$ smearing to match the measured momentum resolution, ensuring the extracted $a_2$ signal is not biased by detector  effects~\cite{ParticleDataGroup:2014cgo}. 
%
To validate the simulation (the same as used in ~\cite{STAR:2023nos,STAR:2023vvb}), templates of individual STARlight physics processes are used to fit $m_{ee}$ and $p_T$ distributions simultaneously with $\chi^2$ minimization. The total template is shown in Fig.~\ref{mass}. The simulated events were selected using
the same track and vertex criteria that were applied to the data. 
The background corrected signal modulation, $a_2^{\rm sig}$,
is a combination of the true $a_2$ and the effects of the detector. Assuming
that the detector effects also produce a nonzero modulation, $a_2^{D} \ne 0$,
the true $a_2$ can be obtained as~\cite{STAR:2022wfe}:
$a_2 = \frac{a_2^{\rm sig} - a_2^{D}}{1-0.5a_2^{\rm sig}a_2^{D}}$.
The $a_2^{D}$ is estimated from the aforementioned simulations.

At low $p_T$, the back-to-back topology of $J/\psi \to e^+e^-$ makes the
reconstructed $\phi$ distribution sensitive to azimuthal variations in acceptance and tracking efficiency, generating a nonzero detector modulation
$a_2^{D}$. The magnitude and $p_T$ dependence of $a_2^{D}$ are determined from full embedding and applied bin-by-bin; following the standard STAR procedure established in azimuthal correlation analyses~\cite{STAR:2024ujm}. More details are provided in the Supplemental Material~\cite{supplemental}.
%
%

Systematic uncertainties are estimated by varying the event, track quality, and electron identification related selection criteria. Simulation parameters related to the $p_T$ reweighting and smearing are also varied to investigate the systematic effects on  $a_2$. Typical systematic uncertainties on the interference signal $a_2$ due to the event, track, and topological selection criteria variations combined are below 6\%, while that due to variations in simulation parameters is estimated to be under 12\%. The data are corrected for TPC tracking efficiency and acceptance effects, with correction residuals included in the systematic uncertainty. 

Figure~\ref{a2} shows the measured interference signal $a_2$ for $J/\psi$ as a function of $p_T$ from Au+Au and isobaric (Ru+Ru \& Zr+Zr) UPCs at $\sqrt{s_{NN}} = 200$ GeV. We compare the measured $J/\psi$ $a_2$ with the published photoproduced $\rho^{0} \rightarrow \pi^{+}\pi^{-}$ results~\cite{STAR:2022wfe} in Au+Au UPCs at $\sqrt{s_{NN}} = 200$ GeV. The interference effects damp out at large $p_T$ where incoherent processes start to dominate. The measured values of $J/\psi$ $a_2$ for $p_T<120$ MeV/$\textit{c}$ are $-0.122 \pm 0.031 \ (\text{stat.}) \pm 0.022 \ (\text{syst.})$ for Au+Au UPCs, and $-0.127 \pm 0.053 \ (\text{stat.}) \pm 0.042 \ (\text{syst.})$ for isobar UPCs. The corresponding significance of negative modulations are $3.2\sigma$ and $1.9\sigma$, respectively. The Au+Au data provide evidence of a negative $\cos(2\phi)$ modulation, opposite in sign to the positive 
modulation reported for $\rho^{0}$. The isobar result has lower statistical significance due to limited statistics, but its negative sign is consistent 
with the Au+Au measurement and provides an independent cross-check of the effect.

The observed sign difference in the $\cos(2\phi)$ modulation between $J/\psi$ and $\rho^{0}$ photoproduction provides direct evidence that the decay-channel spin structure controls the sign of the modulation within a common two-source interference mechanism: $J/\psi\!\to\!e^{+}e^{-}$ (spin-1/2 daughters) yields a negative modulation, whereas $\rho^{0}\!\to\!\pi^{+}\pi^{-}$ (spin-0 daughters) exhibits a positive one ~\cite{Mantysaari:2023prg,Hagiwara:2020juc,Brandenburg:2022jgr}. 
Because the pions are spin-0 bosons, the role of daughter spin is not apparent in the $\rho^{0}$ channel alone. The spin-$1/2$ fermionic decay channel of the $J/\psi$ uniquely reveals the role of daughter spin through the observed sign flip in the modulation amplitude. The peak $J/\psi$ modulation is smaller than that of the $\rho^{0}$, consistent with expectations~\cite{Mantysaari:2023prg} that its more compact dipole reduces the two-source overlap. At maximum, the modulation reaches $-0.220 \pm 0.100$ (stat. and syst. combined) for the isobar systems and $-0.190 \pm 0.077$ (stat. and syst. combined) for Au+Au, about half of the $\rho^{0}$ peak, which is measured to be $0.391 \pm 0.013$ (stat. and syst. combined). 



 %
 %

Theoretical calculations indicate that the smaller radii $(R)$ of Ru and Zr are expected to strengthen the interference effect~\cite{Mantysaari:2023prg}. The $p_T$ dependence of the signal is tied to the impact parameter ($b$), which makes the interference term proportional to $\cos(p_T b)$. Larger  nuclear radii lead to a larger typical $b\ge 2R$, causing the $\cos(p_T b)$ term to oscillate rapidly and suppressing interference at low $p_T$. Consistent with this geometric expectation, the signal strength for $p_{T}<60$ MeV/c appears more pronounced in the isobaric system than in Au+Au collisions (isobar $|a_2|$> Au+Au $|a_2|$).


The data in Fig.~\ref{a2} are compared to STARlight~\cite{Klein:2016yzr}, which assumes no spin interference and therefore predicts $a_2=0$. In contrast, the CGC-based calculations (Model-I)~\cite{Mantysaari:2023prg}, which include linear photon polarization, two-source interference, and finite  photon transverse momentum, reproduce both the sign and approximate magnitude of the measured $a_2$. The agreement is noticeably better for isobars, while  Au+Au data show a different $p_T$ dependence, indicating further model  refinement is needed for larger nuclei. The main Model-I uncertainty comes from the assumed impact-parameter range. 
%
%
The data are also compared to the CGC effective theory Model-II calculations~\cite{Brandenburg:2022jgr}, which extend the framework by incorporating final-state soft-photon radiation. This QED effect generates large positive $a_2$ at higher $p_T$. Our data, which remain near zero in this region, can constrain this model in future studies.




In summary, we see evidence of spin interference in photoproduced $J/\psi$ mesons in ultraperipheral
collisions, exploiting their longer lifetime and delocalized wave function. Unlike $\rho^0$, $J/\psi$ exhibits a negative modulation for $p_T<120$ MeV/$\textit{c}$,  revealing an interference mechanism whose sign is governed by the two fermionic wave functions at a finite separation. As a heavier meson ($3.1$ GeV/$c^2$ vs. $0.77$ GeV/$c^2$ for $\rho^0$), $J/\psi$
probes gluon distributions at finer spatial scales, making it stringent test of QCD effective theories like CGC. 
The smaller $J/\psi$ peak amplitude relative to the $\rho^0$ is consistent with theoretical expectations from its more compact dipole size.
While current CGC calculations qualitatively capture the sign they do not fully reproduce the $p_T$ dependence, highlighting the need for theoretical refinements. 

These results also provide the first direct experimental evidence that polarization-driven angular interference in heavy vector mesons is measurable in photon-induced reactions. This finding directly supports recent proposals to exploit $\cos(2\phi)$ modulations in $J/\psi$ decays as a powerful imaging tool in diffractive $e$+A scattering at the Electron-Ion-Collider~\cite{Kesler:2025ksf}. Together, these developments highlight angular spin interference as a new degree of freedom for resolving gluon spatial structure, beyond what is accessible through $t$-dependent cross sections alone. 


\section*{Acknowledgment}
We thank the RHIC Operations Group and SCDF at BNL, the NERSC Center at LBNL, and the Open Science Grid consortium for providing resources and support.  This work was supported in part by the Office of Nuclear Physics within the U.S. DOE Office of Science, the U.S. National Science Foundation, National Natural Science Foundation of China, Chinese Academy of Science, the Ministry of Science and Technology of China and the Chinese Ministry of Education, NSTC Taipei, the National Research Foundation of Korea, Czech Science Foundation and Ministry of Education, Youth and Sports of the Czech Republic, Hungarian National Research, Development and Innovation Office, New National Excellency Programme of the Hungarian Ministry of Human Capacities, Department of Atomic Energy and Department of Science and Technology of the Government of India, the National Science Centre and WUT ID-UB of Poland, German Bundesministerium f\"ur Bildung, Wissenschaft, Forschung and Technologie (BMBF), Helmholtz Association, Ministry of Education, Culture, Sports, Science, and Technology (MEXT), Japan Society for the Promotion of Science (JSPS), and Agencia Nacional de Investigacion y Desarrollo de Chile (ANID), Chile.

\section{Supplemental Material: Corrections}

\label{appendix}

\subsection{Combinatorial Backgrounds}

The combinatorial background contribution to the measured azimuthal modulation
is determined using the raw (detector-level) invariant-mass dependence of the
modulation coefficient $a_{2}(m_{ee})$. For each $m_{ee}$ interval, the azimuthal
distribution is fitted with
\begin{equation}
f(\phi) = 1 + a_{2}\cos(2\phi).
\end{equation}

\begin{figure}[!t]
  \centering
  \includegraphics[width=0.98\linewidth]{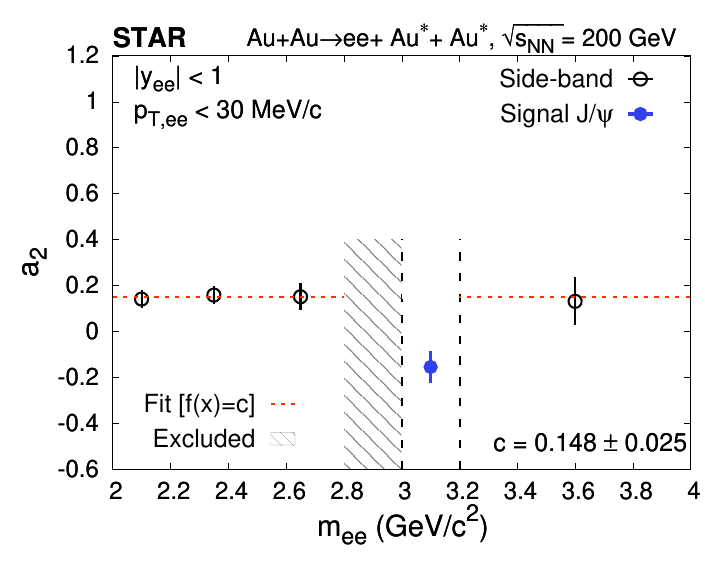}
  \\
 \includegraphics[width=0.98\linewidth]{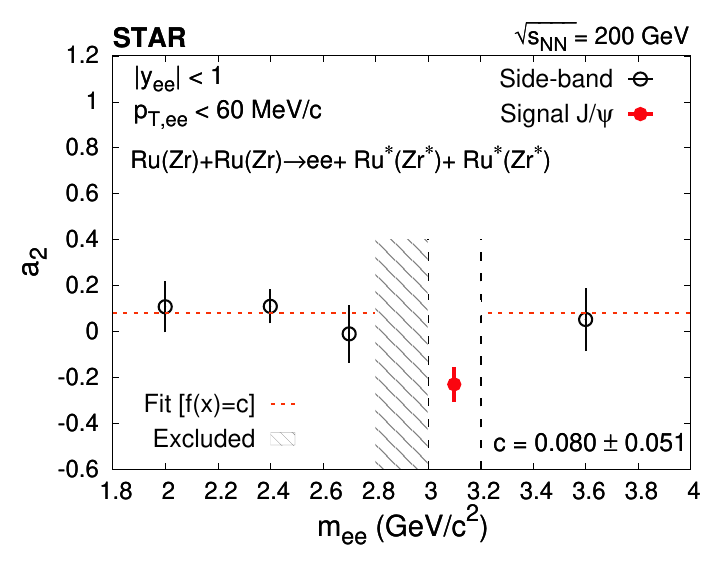}
\caption{
Determination of the background modulation $a_{2}^{\mathrm{bkg}}$ for Au+Au (top) and isobaric (bottom) UPCs at $\sqrt{s_{NN}} = 200~\mathrm{GeV}$ for the indicated $p_{T}$ intervals. Open circles represent the raw $a_{2}(m_{ee})$ values extracted from unlike-sign $e^{+}e^{-}$ sidebands, with the red dashed line showing the constant fit used to define the background level. The $J/\psi$ signal
window (vertical dashed lines) and the radiative-tail region (hatched band) are excluded from the fit.}
\label{fig:a2_bkg}
\end{figure}

To estimate the background modulation ($a_{2}^{\mathrm{bkg}}$) within the
$J/\psi$ region, a constant fit is applied to the $a_{2}(m_{ee})$ values extracted
from unlike-sign $e^{+}e^{-}$ pairs in the sideband mass intervals
$2.0$--$2.8~\mathrm{GeV}/c^{2}$ and $3.2$--$4.0~\mathrm{GeV}/c^{2}$. As shown in
Fig.\ref{fig:a2_bkg}, the signal window ($3.0$--$3.2~\mathrm{GeV}/c^{2}$) and the adjacent
radiative-tail region ($2.8$--$3.0~\mathrm{GeV}/c^{2}$) are excluded from this fit.
This exclusion suppresses contamination from true
$J/\psi \rightarrow e^{+}e^{-}$ candidates that have undergone detector bremsstrahlung.

For the representative low-$p_{T}$ bins illustrated in Fig.\ref{fig:a2_bkg}, the resulting fit
yields $a_{2}^{\mathrm{bkg}} = 0.148 \pm 0.025$ for Au+Au and
$0.080 \pm 0.051$ for the isobar systems. As an independent cross-check of the background determination, we also extract $a_2$ from like-sign $e^{\pm}e^{\pm}$ pairs evaluated within the $J/\psi$ signal window; this provides a background-only control sample at the mass peak.

\begin{figure}[!t]
  \centering
  \includegraphics[width=0.98\linewidth]{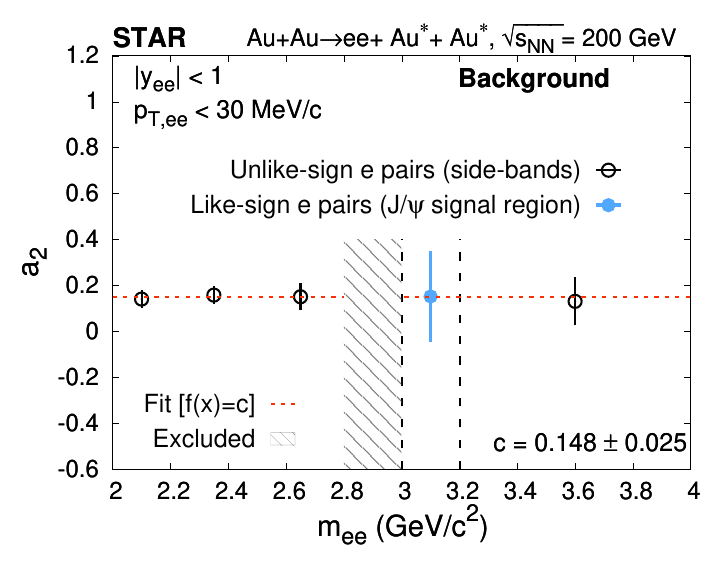}
\caption{ Background modulation $a_{2}^{\mathrm{bkg}}$ for Au+Au UPCs at $\sqrt{s_{NN}}=200$~GeV in the interval $p_{T}<30$~MeV/$c$. Open circles show $a_{2}(m_{ee})$ extracted from unlike-sign $e^{+}e^{-}$ pairs in invariant-mass sideband bins. The blue filled marker shows $a_{2}$ extracted from like-sign $e^{\pm}e^{\pm}$ pairs within the $J/\psi$ signal window, providing a background-only control sample at the mass peak. The red dashed line is a constant fit to the unlike-sign sideband points used to define $a_{2}^{\mathrm{bkg}}$. The $J/\psi$ signal window (vertical dashed lines) and the radiative-tail region (hatched band) are excluded from the sideband fit to suppress signal leakage.}
  \label{fig:a2_ls}
\end{figure}

Figure~\ref{fig:a2_ls} compares the unlike-sign sideband-based background modulation with the like-sign result for Au+Au UPCs at $\sqrt{s_{NN}}=200$~GeV in the bin $p_{T}<30$~MeV/$c$.
 The like-sign value is consistent with the sideband estimate within uncertainties, with a larger uncertainty due to limited like-sign statistics, further supporting the robustness of the background-modulation procedure. 
 

\begin{figure}[htb]
  \centering
  \includegraphics[width=0.98\linewidth]{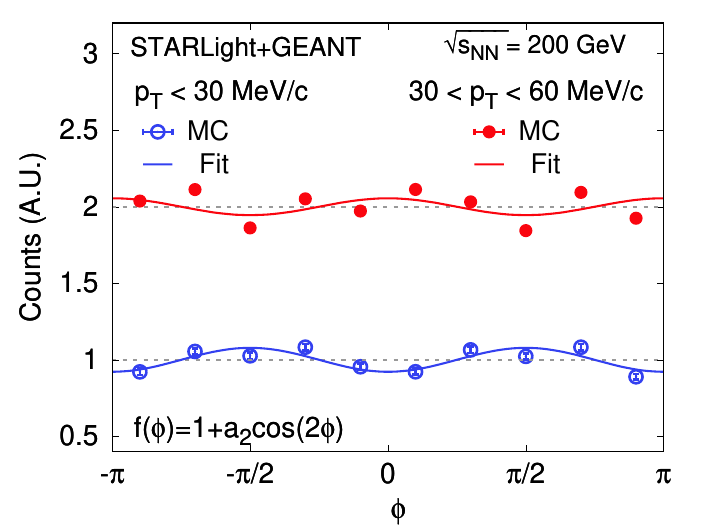}
  \caption{Detector-only $\phi$ modulation from STARlight+GEANT with STARlight input $a_2^{\rm true}=0$. 
  STARlight points are first embedded into zero-bias data, then a full GEANT3-based reconstruction is performed. We refer to this as ``MC''. The curves are fit to the MC points with $1+a_2^{D}\cos(2\phi)$.
  }
  \label{fig:sim_acceptance}
\end{figure}

\subsection{Detector Effect}
To quantify detector-induced modulations in the $\phi$ distribution, we use 
STARlight $J/\psi \to e^{+}e^{-}$ UPC events with no physical 
$\cos(2\phi)$ input ($a_2^{T}=0$). The events are embedded into zero-bias data and processed through the full GEANT3 simulation and STAR reconstruction chain, with identical selections to those used for data. This procedure, widely employed in recent STAR analyses [45], provides a realistic estimate of the detector response.

The reconstructed $\phi$ spectra in each $p_T$ bin are fitted using
\[
N(\phi)=A\,[1+a_2^{D}\cos(2\phi)],
\]
yielding a bin-by-bin determination of $a_2^{D}$. For illustration, 
Fig.~\ref{fig:sim_acceptance} shows the reconstructed distributions for two low-$p_T$ bins.

From these fits we extract $a_2^{D} = -0.0781 \pm 0.0091$ for $p_T < 30$~MeV/$c$ 
and $a_2^{D} = 0.0275 \pm 0.0068$ for $30 < p_T < 60$~MeV/$c$. These values quantify 
the detector response that is removed via the correction formula described in the main 
text. No physical conclusions are drawn from the magnitude of $a_2^{D}$ itself; only the 
corrected observable $a_2^{T}$ is used in the physics interpretation.

\bibliography{main}

\end{document}